\documentstyle[12pt]{article}
\newcommand{\be}{\begin{equation}}
\newcommand{\ee}{\end{equation}}
\newcommand{\ba}{\begin{eqnarray}}
\newcommand{\ea}{\end{eqnarray}}
\newcommand{\baa}{\begin{eqnarray*}}
\newcommand{\eaa}{\end{eqnarray*}}
\newcommand{\bb}{\begin{thebibliography}}
\newcommand{\eb}{\end{thebibliography}}
\newcommand{\ci}[1]{\cite{#1}}
\newcommand{\bi}[1]{\bibitem{#1}}
\begin {document}
\title{ THE EMC SPIN CRISIS: COMPARING PROTON AND PHOTON}
\author{ O.V. TERYAEV\\
{\it Joint Institute for Nuclear Research} \\
{\it 141980, Dubna, Russia}}

\maketitle
\begin{abstract}
The cancellation of photonic and gluonic anomalies naturally explains
the zero singlet contributions to the first moments of
proton and photon spin structure functions.
\end{abstract}
The so-called EMC Spin Crisis,i.e. the small(comparable with zero) value of
the singlet axial current proton matrix element,initiated
a stream of theoretical papers during the
last four years(see,e.g.,\ci{1}). I'll discuss,however,only the
"anomalous way" of its resolution,exploiting the gluonic triangle anomaly
\ci{2}$^{\mbox {\small -}}$\ci{4}.
Due to its natural interpretation as a large contribution of the
 polarized
gluon density via the anomalous part of the box diagram,it is often thought of
\ci{DoKo} as an essentially perturbative QCD approach.
The starting point of the analysis,
mentioned already in \ci{2} is,however,insensitive to any perfect
perturbative approximation.
We need only such quite a general result to be valid
as the expression of the first moment of the spin dependent structure
function $g_1^1$ through the matrix element of the axial current,the latter
 being the only gauge-invariant local operator with the required quantum
  numbers(although either gauge-dependent or nonlocal operators may
  contribute provided some perfect recipe of small and large distances
  separation is adopted). On the other hand, it is not conserved due to the
  axial anomaly. Because of its nonzero anomalous dimension it is a bad
  candidate for being in the angular momentum sum rule. The required
   difference is just the anomalous gluon contribution.

  Therefore,if the
  anomalous contribution is present,the large(i.e. not suppressed
  parametrically) deviation of $g_1^1$ from the naive expectation is naturally
   explained. One may ask, however: is it possible to give a natural
   explanation
   of the zero result? There are several positive answers \ci{F,KZ}. The
   required cancellation occurs,as a rule, after a more or a less complicated
  calculation. Here I'll try to show the zero result to be in a direct
  relationship with  an other remarkable phenomenon occurring in the singlet
  channel, namely, the $U_A(1)$--problem (the absence of the Goldstone meson
  in the zero quark mass limit).

  Such a relation is already known for a few years\cite{EST,2}:
 the ghost pole which provides a mass to $\eta'$ under
 the mixing with the Goldstone $\eta_0'$ generates the additional form
 factor $\tilde G_2$ of the topological gluon current "nearly forward" matrix
 element.
\ba
\langle P , S |K_{\mu}^5(0)| P+q ,S \rangle =2MS_{\mu}\tilde G_1(q^2) +
q_{\mu}(Sq)\tilde G_2(q^2) , \\
  q^{2}\tilde G_2(q^2)|_0 \neq 0.
\ea
One should compare this equation with the gauge invariant axial current
matrix element case:
\ba
\langle P , S |J_{\mu}^5(0)| P+q ,S \rangle =2MS_{\mu}G_1+q_{\mu}(Sq)G_2 , \\
  q^{2}G_2|_0 = 0.
\ea
 The conserved quark--gluon current matrix element (its relation
 to the quark spin is more convenient than the naive one) is expressed
 via the low-energy $\eta_0'$ constants and is comparable with unity.
\be
\Delta \Sigma ={ \sqrt {N_F}\over{2M}} f_{\eta'} g_{\eta_0' NN}.
\ee
This equation shows that the mentioned profound relation between the
Spin Crisis
and $U_A(1)$ problem does exist. However, it is (i) again of the numerical
origin and (ii) doesn't explain the zero EMC result. Moreover, there is an
approach \ci{KiPet}, in which the zero $q$  behaviour of $G_2$ and the
occurrence of massless particles are not related. The main argument is the
finite zero $q$ limit of $q^{\mu}(Sq)/q^2$ provided $P$ and $P+q$ are on
the same mass shell. My first point is to show that it is not the case:
the massless pole is abandoned also by the kinematical arguments.

I start with the following expression for the quark--gluon angular momentum
density
\be
M^{\mu, \nu \rho} = {1\over{2}}\epsilon^{\mu \nu \rho \sigma} J^5_{S\sigma} +
x^{\nu} T^{\mu \rho} - x^{\rho} T^{\mu \nu}.
\ee
The first term in the r.h.s. is just the canonical quark spin tensor. Note
that the energy--momentum tensor here accumulates also the
quark orbital momentum
as the {\it total} gluon angular momentum. We may proceed further along this
way
and express the quark spin in the orbital form with the simultaneous
change of the energy--momentum tensor discovered by Belinfante long ago
\be
M^{\mu, \nu \rho} = x^{\nu} T^{\mu \rho}_B - x^{\rho} T^{\mu \nu}_B.
\ee
As the conservation of the angular momentum
\be
\partial_{\mu} M^{\mu, \nu \rho} = 0
\ee
immediately leads to the symmetry of $T^{\mu \rho}$, the latter implies that
\be
\epsilon_{\mu \nu \rho \alpha} M^{\mu, \nu \rho} = 0.
\ee
One should conclude that the totally antisymmetric quark spin tensor is
somehow
cancelled and doesn't contribute to the total angular momentum \ci{JafMa}.
Is it nevertheless possible to extract any quantitative information about it?

To do this, one may substitute (6) to (7) and (9) and eliminate $T$
(or, more exactly, its antisymmetric part) from the obtained system of
two equations. In the case of
classical fields one obtains:
\be
(g_{\rho \nu} g_{\alpha \mu} - g_{\rho \mu} g_{\alpha \nu})
\partial^{\rho} (J_{5S}^{\alpha} x^{\nu}) = 0.
\ee
It is just the neglect of surface terms necessary to apply the Belinfante
procedure. Passing to the quantum operators one should integrate (6)--(9)
over the whole 3--dimensional space and switch them between nucleon states
with the momenta $P$ and $P + q$. As a result,
${\partial \over{\partial x^{\mu}}}$ is substituted by $-iq_{\mu}$ and
$x^{\mu}$ by $i {\partial \over{q_{\mu}}}$ acting on  $\delta(\vec q)$.
The latter is, by definition, equal, up to sign, to the derivative
acting on the matrix element. One should obtain straightforward,
instead of (10), the following restriction:
\be
q^{\mu}{\partial \over{\partial q^{\alpha}}} |_0
\langle P | J_{5S}^{\alpha}| P+q\rangle =
q^{\alpha}{\partial \over{\partial q^{\alpha}}} |_0
\langle P | J_{5S}^{\mu}| P+q\rangle .
\ee
To make it more clear, let us multiply both sides by $q_{\mu}$.

\begin{equation}
q^2{\partial \over{\partial q^{\alpha}}}
\langle P | J_{5S}^{\alpha}| P+q\rangle
=(q^{\beta}{\partial \over{\partial q^{\beta}}}-1)
q_{\gamma}\langle P | J_{5S}^{\gamma} |P+q \rangle .
\end{equation}

This equality is obviously valid up to the second and higher powers of $q$.
Note that the differential operator in the r.h.s. clearly subtracts the terms
linear in $q$ from the divergence matrix element proportional to
$sq$ for the pure kinematical reasons. The $G_2$ pole term, however,
provides a contribution linear in $q$ to the l.h.s. This contradiction
is manifested already in (10): the $G_2$ pole contribution to the r.h.s.
is zero, while to the l.h.s. is non--zero.

 It is a "kinematical" solution of the $U_A(1)$ problem:
 the special feature of the singlet channel is its relation to the nucleon
 spin,
 rather than existence of the anomaly-induced ghost contribution \ci{DyaEi}.
 The zero mass pole is thus abandoned as both by the kinematical and
  dynamical analysis, in contradiction with \cite{KiPet}.

These deep relations are,however, insufficient to explain why the deviation
from the naive value leads to the zero result.
One should incorporate the dynamics of axial anomaly. I'll try to show
that it is sufficient to suppose the famous t'Hooft consistency principle
\ci{tH} to be valid: the anomalies should be the same for the fundamental
and composite particles.

Let us start with a more clear case of the photon matrix elements. The
axial current on--shell matrix element is described by the single form factor
\be
\langle \gamma(P,\mu) | J_{5}^{\alpha}| \gamma(P+q,\nu)\rangle =
G_{\gamma}(q^2)q^{\alpha}\epsilon^{\mu \nu \rho \sigma} P_{\rho} q_{\sigma} .
\ee
Axial anomaly manifests itself in the zero quark mass limit as the massless
pole in $G_{\gamma}$ \ci{DZ}. At the hadron level the t'Hooft consistency
principle requires either massless baryons (to be inserted into the
triangle loop) or mesons (manifested in the simplest pole diagram). It is
the second possibility which is realized in the Nature. The only exception
is the singlet channel: the mentioned $U_A(1)$ problem leads to the
conservation of the axial current in the chiral limit. Taking into account
gluonic and photonic anomalies one concludes that their contributions
should cancel each other. As a result, one obtains the low--energy QCD theorem
\ci{Sh}:
\be
\langle \gamma(P) | {\alpha\over{2\pi}}F\tilde F| \gamma(P+q)\rangle +
\langle \gamma(P) | {\alpha_s\over{2\pi}}G\tilde G| \gamma(P+q)\rangle = 0.
\ee
I would like to stress that this equality immediately leads to:
\be
\langle \gamma(P,\mu) | J_{5S}^{\alpha}| \gamma(P,\nu)\rangle = 0.
\ee
It may be called the {\it Photon Spin Crisis}. In contrast to the
proton one it seems to be very natural from the kinematical point of view.
The zero $q$ limit of (13) (note that $q^2$ is zero only if all its components
are zero, provided both photons are on--shell) may be non--zero only due to
the anomalous pole. The cancellation of the gluonic and photonic anomalies
in the singlet channel just restores the naive result. As to the non--singlet
channel the anomalous pole provides the calculable values of the forward
axial current matrix element. As a result, one obtains the sum rule for
the photon spin structure function $g_1^{\gamma}$ \ci{ET-Ph}:
\be
\int_0^1 g_1^{\gamma}(x) dx = - N_C{\langle e^4 \rangle\over{2}}
{\alpha\over{2\pi}}.
\ee

It has been recently criticised \ci{Nar} for the following two main reasons.

First, for the on--shell photons and massive quarks the r.h.s. of the sum rule
is zero. I would like to mention that it was stated already in \ci{ET-Ph}
(see eq.(4) of this Ref.), where the reason for the
zero result, namely, the cancellation
of the "normal" and anomalous contributions, was identified. It seems that
to observe the anomalous contribution the photon virtuality should exceed
$m_{\pi}$ (not $m_{\rho}$, as stated in \ci{Nar}) that replaces $m_q$ when
confinement is taken into account. As it is well known, the $m_\pi$ is
proportional to $m_q$ and turns to zero simultaneously with it, while
$m_\rho$ is determined by the condensates $\langle G^2\rangle$ and
$\langle \bar q q \rangle^2$ and is independent of the light quark mass.
The expected virtuality dependence
was discussed in the case of the conformal anomaly contribution to
$\gamma \gamma$--scattering \ci{GorIo}. The anomalous contribution can
also be isolated making the transverse momentum cutoff, analogously
to the anomalous gluon contribution in the proton case \ci{4}.
The latter method seems to be more realistic for the experimental
verification of the sum rule.

Second, the specific feature of the singlet channel was missed in \ci{ET-Ph}.
However, the ratio of the meson coupling constants for the singlet and
non--singlet channels, appearing in the sum rule r.h.s. \ci{Nar},
may be eliminated. The t'Hooft consistency principle and the cancellation
of the gluon and photon anomalies lead to the zero singlet channel
contribution which results in the sum rule:
\be
\int_0^1 g_1^{\gamma}(x) dx = - N_C{\langle e^4 \rangle\over{2}}
{\alpha\over{2\pi}}(1 - c)
\ee
Here $c=\langle e^2 \rangle^2/(N_F \langle e^4 \rangle)$
is the ratio of singlet
and nonsinglet weights \ci{Nar}. Note that the zero singlet channel
contribution does not imply the zero decay width of $\eta_0^{'}\rightarrow
\gamma \gamma$
because of the ghost pole which substitute the anomalous one.

Passing to the hadron matrix element case one should confront the problem
of dealing with {\it two} above mentioned form factors instead of one.
Although the situation here is not yet clear enough, it is possible to
give the simple diagrammatic explanation of the EMC result.

The Photon Spin Crisis means that the total transition amplitude of the singlet
axial current to two photons is zero
provided the direct annihilation $q\bar q\rightarrow \gamma \gamma$ (Fig. 1a)
and  that via a gluon pair (Fig. 1b) are taken into account.

\begin{figure}
\hspace{15pt}
\unitlength=2.50pt
\special{em:linewidth 0.4pt}
\linethickness{0.4pt}
\begin{picture}(134.00,80.00)
\put(120.00,40.95){\oval(28.00,14.00)[]}
\put(129.00,65.00){\line(-1,0){18.00}}
\put(111.00,65.00){\line(0,0){0.00}}
\put(120.00,80.00){\line(-3,-5){9.00}}
\put(129.00,65.00){\line(-3,5){9.00}}
\put(41.00,65.00){\line(-1,0){18.00}}
\put(32.00,80.00){\line(-3,-5){9.00}}
\put(41.00,65.00){\line(-3,5){9.00}}
\put(110.53,63.50){\oval(3.01,3.01)[l]}
\put(110.53,61.49){\oval(3.01,3.01)[l]}
\put(111.03,62.49){\oval(2.01,1.00)[r]}
\put(111.03,60.49){\oval(2.01,1.00)[r]}
\put(110.53,59.48){\oval(3.01,3.01)[l]}
\put(110.53,57.47){\oval(3.01,3.01)[l]}
\put(111.03,58.48){\oval(2.01,1.00)[r]}
\put(110.53,55.46){\oval(3.01,3.01)[l]}
\put(110.53,53.46){\oval(3.01,3.01)[l]}
\put(111.03,54.46){\oval(2.01,1.00)[r]}
\put(111.03,52.45){\oval(2.01,1.00)[r]}
\put(110.53,51.45){\oval(3.01,3.01)[l]}
\put(110.53,49.44){\oval(3.01,3.01)[l]}
\put(111.03,50.44){\oval(2.01,1.00)[r]}
\put(111.03,56.47){\oval(2.01,1.00)[r]}
\put(129.03,65.00){\line(0,0){0.00}}
\put(128.56,63.50){\oval(3.01,3.01)[l]}
\put(128.56,61.49){\oval(3.01,3.01)[l]}
\put(129.06,62.49){\oval(2.01,1.00)[r]}
\put(129.06,60.49){\oval(2.01,1.00)[r]}
\put(128.56,59.48){\oval(3.01,3.01)[l]}
\put(128.56,57.47){\oval(3.01,3.01)[l]}
\put(129.06,58.48){\oval(2.01,1.00)[r]}
\put(128.56,55.46){\oval(3.01,3.01)[l]}
\put(128.56,53.46){\oval(3.01,3.01)[l]}
\put(129.06,54.46){\oval(2.01,1.00)[r]}
\put(129.06,52.45){\oval(2.01,1.00)[r]}
\put(128.56,51.45){\oval(3.01,3.01)[l]}
\put(128.56,49.44){\oval(3.01,3.01)[l]}
\put(129.06,50.44){\oval(2.01,1.00)[r]}
\put(129.06,56.47){\oval(2.01,1.00)[r]}
\put(40.97,63.96){\oval(2.01,2.01)[l]}
\put(40.97,62.00){\oval(2.01,1.91)[r]}
\put(40.97,60.04){\oval(2.01,2.01)[l]}
\put(40.97,58.03){\oval(2.01,2.01)[r]}
\put(40.97,56.01){\oval(2.01,2.01)[l]}
\put(40.97,54.05){\oval(2.01,1.91)[r]}
\put(40.97,52.09){\oval(2.01,2.01)[l]}
\put(40.97,50.08){\oval(2.01,2.01)[r]}
\put(40.97,48.06){\oval(2.01,2.01)[l]}
\put(40.97,46.05){\oval(2.01,2.01)[r]}
\put(23.04,63.96){\oval(2.01,2.01)[l]}
\put(23.04,62.00){\oval(2.01,1.91)[r]}
\put(23.04,60.04){\oval(2.01,2.01)[l]}
\put(23.04,58.03){\oval(2.01,2.01)[r]}
\put(23.04,56.01){\oval(2.01,2.01)[l]}
\put(23.04,54.05){\oval(2.01,1.91)[r]}
\put(23.04,52.09){\oval(2.01,2.01)[l]}
\put(23.04,50.08){\oval(2.01,2.01)[r]}
\put(23.04,48.06){\oval(2.01,2.01)[l]}
\put(23.04,46.05){\oval(2.01,2.01)[r]}
\put(129.00,34.00){\line(-1,0){18.00}}
\put(128.97,32.96){\oval(2.01,2.01)[l]}
\put(128.97,31.00){\oval(2.01,1.91)[r]}
\put(128.97,29.04){\oval(2.01,2.01)[l]}
\put(128.97,27.03){\oval(2.01,2.01)[r]}
\put(128.97,25.01){\oval(2.01,2.01)[l]}
\put(128.97,23.05){\oval(2.01,1.91)[r]}
\put(128.97,21.09){\oval(2.01,2.01)[l]}
\put(128.97,19.08){\oval(2.01,2.01)[r]}
\put(128.97,17.06){\oval(2.01,2.01)[l]}
\put(128.97,15.05){\oval(2.01,2.01)[r]}
\put(111.04,32.96){\oval(2.01,2.01)[l]}
\put(111.04,31.00){\oval(2.01,1.91)[r]}
\put(111.04,29.04){\oval(2.01,2.01)[l]}
\put(111.04,27.03){\oval(2.01,2.01)[r]}
\put(111.04,25.01){\oval(2.01,2.01)[l]}
\put(111.04,23.05){\oval(2.01,1.91)[r]}
\put(111.04,21.09){\oval(2.01,2.01)[l]}
\put(111.04,19.08){\oval(2.01,2.01)[r]}
\put(111.04,17.06){\oval(2.01,2.01)[l]}
\put(111.04,15.05){\oval(2.01,2.01)[r]}
\put(29.00,3.00){\makebox(0,0)[cc]{a)}}
\put(119.00,4.00){\makebox(0,0)[cc]{b)}}
\end{picture}
\caption{ Cancellation of the photonic (a) and gluonic (b) anomalies of
the singlet axial current}
\end{figure}
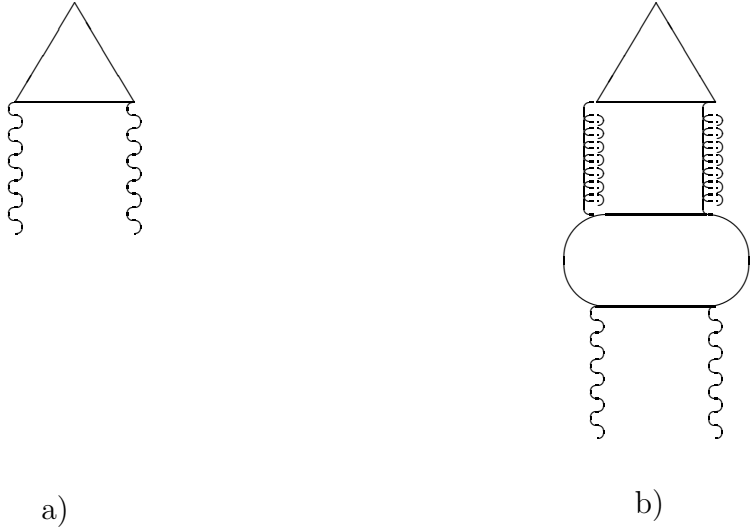

Note that the anomaly in the dispersion \ci{DZ} approach is generated by the
horizontal quark propagator whose "dynamical" smallness $(O(m^{-2}))$ for the
forward scattering compensates the "kinematical" smallness of the nominator
$(O(m^2))$ due to the chirality breaking. The axial current vertex is then
just the electron--positron source with the required parity. Let us therefore
"cut off"\ci{OT91} the upper part of Fig. 1, passing to Fig.2.

\begin{figure}
\hspace{15pt}
\unitlength=2.50pt
\special{em:linewidth 0.4pt}
\linethickness{0.4pt}
\begin{picture}(143.00,80.00)
\put(120.00,41.95){\oval(28.00,14.00)[]}
\put(110.53,64.50){\oval(3.01,3.01)[l]}
\put(110.53,62.49){\oval(3.01,3.01)[l]}
\put(111.03,63.49){\oval(2.01,1.00)[r]}
\put(111.03,61.49){\oval(2.01,1.00)[r]}
\put(110.53,60.48){\oval(3.01,3.01)[l]}
\put(110.53,58.47){\oval(3.01,3.01)[l]}
\put(111.03,59.48){\oval(2.01,1.00)[r]}
\put(110.53,56.46){\oval(3.01,3.01)[l]}
\put(110.53,54.46){\oval(3.01,3.01)[l]}
\put(111.03,55.46){\oval(2.01,1.00)[r]}
\put(111.03,53.45){\oval(2.01,1.00)[r]}
\put(110.53,52.45){\oval(3.01,3.01)[l]}
\put(110.53,50.44){\oval(3.01,3.01)[l]}
\put(111.03,51.44){\oval(2.01,1.00)[r]}
\put(111.03,57.47){\oval(2.01,1.00)[r]}
\put(128.56,64.50){\oval(3.01,3.01)[l]}
\put(128.56,62.49){\oval(3.01,3.01)[l]}
\put(129.06,63.49){\oval(2.01,1.00)[r]}
\put(129.06,61.49){\oval(2.01,1.00)[r]}
\put(128.56,60.48){\oval(3.01,3.01)[l]}
\put(128.56,58.47){\oval(3.01,3.01)[l]}
\put(129.06,59.48){\oval(2.01,1.00)[r]}
\put(128.56,56.46){\oval(3.01,3.01)[l]}
\put(128.56,54.46){\oval(3.01,3.01)[l]}
\put(129.06,55.46){\oval(2.01,1.00)[r]}
\put(129.06,53.45){\oval(2.01,1.00)[r]}
\put(128.56,52.45){\oval(3.01,3.01)[l]}
\put(128.56,50.44){\oval(3.01,3.01)[l]}
\put(129.06,51.44){\oval(2.01,1.00)[r]}
\put(129.06,57.47){\oval(2.01,1.00)[r]}
\put(40.97,56.96){\oval(2.01,2.01)[l]}
\put(40.97,55.00){\oval(2.01,1.91)[r]}
\put(40.97,53.04){\oval(2.01,2.01)[l]}
\put(40.97,51.03){\oval(2.01,2.01)[r]}
\put(40.97,49.01){\oval(2.01,2.01)[l]}
\put(40.97,47.05){\oval(2.01,1.91)[r]}
\put(40.97,45.09){\oval(2.01,2.01)[l]}
\put(40.97,43.08){\oval(2.01,2.01)[r]}
\put(40.97,41.06){\oval(2.01,2.01)[l]}
\put(40.97,39.05){\oval(2.01,2.01)[r]}
\put(23.04,56.96){\oval(2.01,2.01)[l]}
\put(23.04,55.00){\oval(2.01,1.91)[r]}
\put(23.04,53.04){\oval(2.01,2.01)[l]}
\put(23.04,51.03){\oval(2.01,2.01)[r]}
\put(23.04,49.01){\oval(2.01,2.01)[l]}
\put(23.04,47.05){\oval(2.01,1.91)[r]}
\put(23.04,45.09){\oval(2.01,2.01)[l]}
\put(23.04,43.08){\oval(2.01,2.01)[r]}
\put(23.04,41.06){\oval(2.01,2.01)[l]}
\put(23.04,39.05){\oval(2.01,2.01)[r]}
\put(129.00,35.00){\line(-1,0){18.00}}
\put(128.97,33.96){\oval(2.01,2.01)[l]}
\put(128.97,32.00){\oval(2.01,1.91)[r]}
\put(128.97,30.04){\oval(2.01,2.01)[l]}
\put(128.97,28.03){\oval(2.01,2.01)[r]}
\put(128.97,26.01){\oval(2.01,2.01)[l]}
\put(128.97,24.05){\oval(2.01,1.91)[r]}
\put(128.97,22.09){\oval(2.01,2.01)[l]}
\put(128.97,20.08){\oval(2.01,2.01)[r]}
\put(128.97,18.06){\oval(2.01,2.01)[l]}
\put(128.97,16.05){\oval(2.01,2.01)[r]}
\put(111.04,33.96){\oval(2.01,2.01)[l]}
\put(111.04,32.00){\oval(2.01,1.91)[r]}
\put(111.04,30.04){\oval(2.01,2.01)[l]}
\put(111.04,28.03){\oval(2.01,2.01)[r]}
\put(111.04,26.01){\oval(2.01,2.01)[l]}
\put(111.04,24.05){\oval(2.01,1.91)[r]}
\put(111.04,22.09){\oval(2.01,2.01)[l]}
\put(111.04,20.08){\oval(2.01,2.01)[r]}
\put(111.04,18.06){\oval(2.01,2.01)[l]}
\put(111.04,16.05){\oval(2.01,2.01)[r]}
\put(29.00,4.00){\makebox(0,0)[cc]{a)}}
\put(119.00,5.00){\makebox(0,0)[cc]{b)}}
\put(110.00,66.00){\line(1,0){19.00}}
\put(129.00,66.00){\line(0,0){0.00}}
\put(110.00,66.00){\line(1,0){1.00}}
\put(111.04,65.94){\line(0,1){8.13}}
\put(129.04,74.06){\line(0,-1){8.13}}
\put(22.00,58.00){\line(1,0){19.00}}
\put(41.00,58.00){\line(0,0){0.00}}
\put(22.00,58.00){\line(1,0){1.00}}
\put(23.04,57.94){\line(0,1){8.13}}
\put(41.04,66.06){\line(0,-1){8.13}}
\put(97.00,60.00){\dashbox{3.00}(46.00,20.00)[cc]{}}
\put(12.00,62.00){\dashbox{3.00}(42.00,18.00)[cc]{}}
\end{picture}
\caption{ Cancellation of the quark (a) and gluon (b) contributions
to the nucleon spin structure function first moment}
\end{figure}
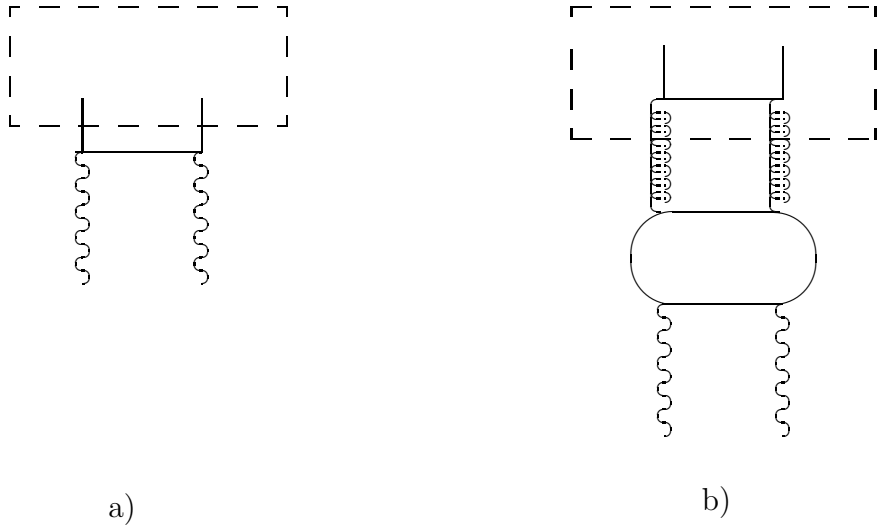

 Note that Fig. 2a
manifests just the quark contribution to the proton polarized DIS while
Fig. 2b the anomalous gluon one (the dashed box represents the hadron--parton
transition). The contributions of Fig. 2a and 2b should cancel just like that
of Fig. 1a and 1b.
It is the diagrammatic explanation of the EMC Proton Spin Crisis.

The photon--gluon blob (Fig. 1b,2b) should be of nonperturbative nature to
cancel $\alpha_s$ additional to Fig. 1a,2a. It was studied in the framework
of the $1/N_C$--expansion \ci{GhPa}, where the validity of the Leutwyler-
Shifman theorem, Fig.1, was checked. Note, however, that this
result has a perturbative counterpart. The running coupling constant
suppression in the asymptotically free gauge theory means, in practice,
the logarithmic decrease in comparison with the Born term, provided the
energy scale tends to infinity. This suppression is absent for the anomalous
gluon contribution \ci{2}, as soon as the Leading Log
corrections (i.e. the ladder
diagrams) are taken into account. The corresponding anomalous dimension
is positive and equal to the one--loop beta--function, compensating the
leading order $\alpha_s$ evolution. The partonic interpretation is the
$\alpha_s^{-1}$ (nonperturbative--like) growth of the gluon distribution
first moment. The reason for this unusual perturbative behaviour is just
the Adler--Bardeen non--renormalization theorem. As the nonperturbative
effects are also generated by the Axial Anomaly, the coincidence of
perturbative and nonperturbative results is just the manifestation of the
t'Hooft consistency principle.

The relation of the simplest quark handbag diagram, Fig 2a, to the
electromagnetic axial anomaly is probably the most unexpected result.
To understand it, let us write down the contribution of Fig. 2a
in the following way:
\be
T^{\mu \nu} \sim
m\epsilon^{\mu \nu \alpha \beta} s_{\alpha}q_{\beta}{1\over{q^2 + 2(Pq)}} = \\
\\\xi_{\parallel}
\epsilon_{\perp}^{\mu \nu} {(Pq)\over{q^2 + 2(Pq)}} = \\
\xi_{\parallel}
\epsilon_{\perp}^{\mu \nu} {\omega\over{2(1 - \omega)}}.
\ee
Here $m$ is the mass, $s$ the covariant and $\xi_{\parallel}$ the longitudinal
polarization of the struck quark,
$\epsilon_{\perp}$ is a two--dimensional antisymmetric tensor in the
hyperplane orthogonal to $P$ and $q$. Expanding this expression in the
unphysical region in powers of $\omega$ one notices
that the constant term is absent.
The same is of course true for the crossed antiquark diagram.
Therefore, the Fig 2a
contribution to the first moment of $g_1$ structure function
is zero. This fact is usually conspired, because normally the expansion
is performed {\it after} extracting the kinematical factor in the nominator.
However, it should be taken into account in the free--field theory:
it arises due to the boost in the quark (not hadron!) momentum direction which
removes $m$ from the first line in (18).

The anomaly manifests itself if one put $q^2$ to zero and immediately obtain
the $\omega$--independent constant. It cannot be extracted directly in the
cross--section because the crucial pole in the quark propagator is changed
to $\delta$--function when the discontinuity is taken. The situation is
analogous to the QED low--energy theorems case: the amplitude is determined
by the pole of the intermediate state propagator and its integral is expressed
via the square of the elastic form factor.
Therefore we expect that the intensively
discussed Gerasimov--Drell--Hearn (GDH) sum rule (see e.g.
\ci{Sof} and ref. therein) in the massless fermion case is
closely related to the axial anomaly.
The zero value is then expected for the singlet contribution to this sum
rule in the close analog with the Photon and Proton Spin Crisis.
This hypothesis is strongly supported by the fact that the transition from the
Bjorken to GDH sum rule is smooth in the nonsinglet channel \ci{Sof}:
no dramatic sign change is observed. It is interesting that for the second
nonsinglet $SU(3)$ combination both Bjorken and GDH sum rules change sign
preserving the possibility of smooth interpolation. However, the transition
$Q_0$ is about 4 times lower in this case. It is possibly due to the fact
that the additional $\langle m q \bar q \rangle$ quark condensates should
be taken into account. Whereas the light quark condensate $\langle q \bar q
\rangle^2$ leads to $G_A/G_V\geq 1$ \ci{Io},
the light and strange quarks contribution should partially cancel each other
to obtain
the experimental value of $G_A/G_V$ in this channel. The perturbative
result may be continued then to the lower $Q$ region.

There is another possibility to obtain a smooth transition to $Q^2=0$,
realized recently
\ci{SoTe}. If one decompose the GDH sum rule into the contributions
of tensor structures $\epsilon^{\mu \nu \alpha \beta} s_\alpha q_\beta$
and $(sq)\epsilon^{\mu \nu \alpha \beta} P_\alpha q_\beta$ (known
as Schwinger sum rules) the first one approach zero smoothly.
The strong $Q^2$ dependence is then described by the elastic contribution
to the second Schwinger sum rule which is nothing else than the known
Burkhardt--Cottingham sum rule for the $G_2$ structure function.
This leads to the zero longitudinal lepton--proton asymmetry at
$Q^2 \sim 0.2 GeV^2$.

This approach is qualitatively supported by the fact that the $\Delta(1232)$
saturating about 80\% of the GDH integral contribute {\it only} via the
second $G_2$ tensor if only the dominant magnetic dipole transition form
factor $G_M$ is nonzero. The contribution to the first tensor arises only via
the interference of the Coulomb quadrupole form factor $G_C$ with the
electric quadruple one $G_E$ and $G_M$.

One may therefore expect that $G_2$ is somehow related to the gluon
anomaly, as well as the $G_M$ of the $\Delta(1232)$. It is not very
strange because both these structures are produced by the helicity-flip,
the only source in massless QCD being just the axial anomaly.
I would like also to compare the EMC and GDH integrals: both are positive
in a chiral limit and are reduced to zero (EMC) and negative value (GDH).
The identification of the anomaly in the later case, however, requires
further investigation.

I conclude with the statement
that the EMC Spin crisis seems to be a natural consequence
of the t'Hooft consistency principle realized via the cancellation
of the QED and QCD anomalies. The problem, however, requires further
investigation, both theoretical and experimental. The latter includes e.g.
search for the Photon Spin Crisis, the neutron spin structure
investigation and the GDH sum rule $Q^2$-dependence.

\bb{99}
\bi{1} A.V.Efremov, Preprint CERN-TH.6466/92.
\bi{2} a) A.V.Efremov, O.V.Teryaev, Preprint JINR E2-88-287.\\
b) A.V.Efremov, J.Soffer, O.V.Teryaev, Nucl.Phys. {\bf B346}, 97(1990).
\bi{3} G.Altarelli, G.G.Ross, Phys.Lett. {\bf B214}, 381(1988).
\bi{4} R.D.Carlitz, J.C.Collins, A.H.Mueller, Phys.Lett. {\bf B214},
229(1988).
\bi{DoKo} A.E.Dorokhov, N.I.Kochelev, Yu.A.Zubov, Argonne Preprint
PHY-7056-TH-92(1992).
\bi{F} S.Forte, E.V.Shuryak, Nucl.Phys. {\bf B357}, 153(1991).
\bi{KZ} J.H.Kuhn, V.I.Zakharov, Phys.Lett. {\bf B252}, 615(1990).
\bi{EST} A.V.Efremov, J.Soffer, N.T\"ornqvist, Phys.Rev. {\bf D44},
1495(1991).
\bi{KiPet} A.V.Kisselev, V.A.Petrov, Preprint CERN-TH.6355/91.
\bi{JafMa} R.L.Jaffe, A.Manohar, Nucl.Phys. {\bf B337}, 509(1990).
\bi{DyaEi} D.I.Dyakonov, M.I.Eides, ZhETF {\bf 81}, 434 (1981).
\bi{tH} G.t'Hooft in {\it Recent Development in Gauge Theories},
New York, Plenum Press, 1980.
\bi{DZ} a) A.D.Dolgov, V.I.Zakharov, Nucl.Phys.{\bf B27}, 525(1971).\\
b) A.D.Dolgov, V.I.Zakharov, I.B.Khriplovich, Pis'ma ZhETF {\bf 45},
   511(1987).
\bi{Sh} H.Leutwyler, M.Shifman, Bern University Preprint BUTP-89/2(1989);\\
M.A.Shifman, Uspekhi Fiz. Nauk {\bf 157}, 561(1989).
\bi{ET-Ph} A.V.Efremov, O.V.Teryaev, Phys.Lett. {\bf B240}, 200(1990).
\bi{Nar}  S.Narison, G.M.Shore, G.Veneziano, Preprint CERN-TH.6467/92.
\bi{GorIo} A.S.Gorski, B.L.Ioffe, A.Yu.Khodjamirian, Phys.Lett. {\bf B227},
474(1989).
\bi{OT91} O.V.Teryaev, ICTP Preprint IC/91/390(1991), JINR Preprint E2-93-25.
\bi{GhPa} K.Ghosh, B.Patel, Columbia University Preprint CU-TP-509(1991).
\bi{Sof}  B.L.Ioffe, V.A.Khoze, L.N.Lipatov, {\it Hard Processes},
North-Holland, 1984.\\ J.Soffer, Preprint CPT-90/P.2443(1990).
\bi{Io}   V.M.Belyaev, I.I.Kogan, Pis'ma ZhETF {\bf 37}, 611(1983).
\bi{SoTe} J.Soffer, O.Teryaev, Preprint CPT-92/P.2836(1992).
\eb
\end{document}